\documentclass{PoS}

\title{\begin{center}
Nuclear properties of Brightest Cluster Galaxies: results and new observations for two peculiar cases. 
\end{center}}

\ShortTitle{Parsec-scale Properties of Brightest Cluster Galaxies}

\author{\speaker{Elisabetta Liuzzo}$^a$, Gabriele Giovannini $^{a}$$^{b}$,
and Marcello Giroletti $^{a}$\\
\llap{$^a$}Istituto di Radioastronomia
 Via P.Gobetti 101, 40129 Bologna, Italy\\
\llap{$^b$}Dipartimento di Astronomia, Universit\`a di Bologna
Via Ranzani 1, 40127 Bologna, Italy\\
E-mail: \email{liuzzo@ira.inaf.it}, \email{ggiovannin@ira.inaf.it}, \email{giroletti@ira.inaf.it}}

\abstract{We present here our results on a complete sample of
  Brightest Cluster Galaxies (BCGs) in nearby Abell Clusters (distance
  class $<$3).  Combined with data from the literature, we provide
  parsec scale information for 34 BCGs. We found that also radio loud BCGs 
have core structures very complex (e.g. 4C 26.42 in Abell 1795). Moreover, we noted a possible
  dichotomy between BCGs in cool-core clusters and those in non-cool-core clusters. Among resolved sources, those in cool-core clusters tend to have two-sided parsec scale jets, while those in less
  relaxed clusters have predominantly one-sided parsec scale jets. We
  suggest that this difference is caused by the interplay
  between the jets and the surrounding medium. Evidence of
  recurrent activity is also found in BCGs in cool-core clusters. For two peculiar cases of BCGs (IC 712 in Abell 1314 and NGC 6047 in Abell 2151) we asked and obtained new VLBA observations a 1.6 GHz. We discuss them here for the first time.}

\FullConference{10th European VLBI Network Symposium and EVN Users Meeting: VLBI and the new generation of radio arrays\\
		September 20-24, 2010\\
		Manchester Uk}

\begin{document}

\section{Introduction}

Brightest Cluster Galaxies (BCGs) are a unique class of objects
\cite{ye04}. These galaxies are the most luminous and massive
galaxies in the Universe, being up to ten times brighter than typical
elliptical galaxies, and having large characteristic radii (tens of kiloparsecs).
 Most BCGs are cD galaxies with extended envelopes
over hundreds of kiloparsecs, but they can also be giant E and D galaxies. 
The optical morphology often shows evidence of past or recent galaxy
mergers, such as multiple nuclei.  They tend to lie very
close to peaks of the cluster X-ray emission and to have velocities
close to the cluster rest frame velocity. All these properties
indicate that they might have experienced an unusual formation history
compared to other E galaxies. Recent models suggest that
BCGs must form earlier, and that galaxy merging within the
cluster, during collapse within a cosmological hierarchy, is a viable
scenario for their formation \cite{ber06}.  In many cool core clusters BCGs often have
blue excess light indicative of recent star formation
\cite{mcn04}. 

In the radio band, it is well established that BCGs have peculiar
properties. They are more likely to be radio-loud than
other galaxies of the same mass \cite{bes06}. Their
radio morphology often also shows evidence of a strong interaction with the
surrounding medium: some BCGs have a wide angle tail structure (WAT)
that is either extended on the kiloparsec scale (e.g., 3C\,465 in A2634),
or of small size (e.g. NGC4874 in Coma cluster); others are diffuse and amorphous sources, either extended
(3C\,84 in Perseus) or of small size (e.g., the BCG
in A154). These last two sources are rare in the general
radio population, but frequently present in BCGs and in particular in
BCGs located in cooling core clusters of galaxies.

Radio-loud AGN in BCGs have been proposed as a potential solution to
the cooling-flow problem \cite{mcn05}. As a consequence of radiative
cooling, the temperature in the central regions of the cluster is
expected to drop. However, XMM-Newton and Chandra observations of
cooling core clusters have shown that the temperature of cluster cores
is $\sim$30$\%$ higher than expected and also the amount of cooling
gas is only about 10$\%$ of that predicted. X-ray cavities in the emitting gas coincident with radio lobes demonstrate the interplay between the radio activity of
BCGs and the slowing or arrest of cooling in cluster centers
\cite{bi08}. Very deep Chandra observations of the Perseus and
Virgo clusters detected
approximately spherical pressure waves in these clusters. These
``ripples'' are excited by the expanding radio bubbles and the
dissipation of their energy can provide a quasi-continuous heating of
the X-ray emitting gas \cite{ru04}.

Despite these results, many important properties of BCGs are poorly
known. Not all BCGs are strong radio sources and cyclic activity with
a moderate duty cycle is necessary to justify the deceleration of the
cooling processes in clusters where the BCG is not a high power radio
source.  Moreover, it is unclear if radio properties of BCGs in
cooling flow clusters systematically differ from those of BCGs
in merging clusters. We note that in cooling clusters the kiloparsec scale
morphology of BCGs is often classified as a mini-halo \cite{gov09}, but extended `normal' sources are also present (e.g., Hydra
A), while in merging clusters the most common
morphology is a WAT source, but point-like and core-halo
sources are also present.

On the parsec scale, BCGs have been poorly studied  as a
class, data being available only for the mostly famous, bright
radio galaxies.  In some of these cases, these BCGs resemble to normal FRI
radio galaxies with relativistic collimated jets. Parsec scale jets
are usually one-sided because of Doppler boosting effects (e.g.,  3C\,465
in A2634 and 0836+29 in A690), although there are also
cases where two-sided symmetric jets are detected in VLBI images, and
the presence of highly relativistic jets is uncertain (e.g. 3C\,338 in
A2199, and Hydra A in A780).

\section{The sample and results.} \label{res}

To properly investigate their properties, we began a statistically study of BGCs on parsec scale.
To search for links between the parsec scale properties of BCGs in rich clusters of galaxies 
and the cluster properties, we identified a sample of clusters with no 
selection effects in terms of X-ray and radio properties, and 
at limited distance for obvious sensitivity and angular resolution reasons.

We therefore defined an unbiased complete sample by selecting all BCGs in nearby
(Distance Class lower than 3) Abell clusters with a Declination greater
than 0$^{\circ}$. All clusters have been included with no selection on
the cluster conditions (e.g., cooling) and no selection on the BCG
radio power. For all these sources, we collected VLBA observations at 6 cm. This implied new observations in 2008 for most of them.

To improve our statistics, we performed a search in the literature and archive data looking for VLBI
data of BCGs in Abell clusters with DC $>$2. We added to our complete sample the following clusters: A690, A780,
A1795, A2052, A2390, A2597 and 3526 (expanded sample). 

The complete sample, the expanded one, the used VLBA data and some preliminary results were yet presented in \cite{liu08}.
We remind here that a strong dichotomy on pc scale morphology between BGCs in cool core and non cool core clusters was found (Table 1). 
Undetected and unresolved sources are found independently to the cluster type. We also noted that in a few objects, episodic jet activity from the central engine of AGN are found . Recurrent activity of radio source in cool core clusters is of great interest to the study of AGN feedback in clusters (\cite{liu10}).

\begin{table}[htp]
\caption{{\bf BCG counts in the expanded sample}.Note that most of undetected sources in VLBA observations are 
in BCG that   are radio quiet (or faint) also in VLA observations.}
\begin{center}
\label{tab:id}
\tabcolsep2mm
\begin{tabular}{ccccccc}
\hline
\hline
Sample & Cluster  & Number & two-sided & one-sided & point & N.D. \\
Expanded & cool core   &  10    & 7 (70$\%$)   & --  & 1  & 2  \\
         & non cool core  &  24   & --  & 14 (58$\%$) & 1  & 9  \\
\hline
\end{tabular}
\end{center}
\end{table}

We compared our results with the Bologna Complete Sample (BCS)
\cite{gio05, liu10} which is 1) a complete sample of 95 FRI radio galaxies
spanning the same radio power range as our BCG sample, 2) free of selection effects in term of jet velocity and
orientation and 3) for which VLBI observations at 5 GHz and kpc morphology information for most of these objects are
available and presented in \cite{liu10}. Among the results from the
BCS study, we found that the one-sided jet
morphology is the predominant structure and only 22$\%$ of FRI radio
galaxies have two-sided jets. This agrees with expectations
based on a random orientation for sources with relativistic jets.
Based on conclusions for the BCS sample, we suggest that all FRIs 
outside cool cores have
similar parsec scale properties regardless of their host galaxy
classification (BCG or non BCG).  One-sided structures in non cool
core clusters are produce by to Doppler boosting effects in relativistic,
intrinsically symmetric jets.

Two-sided structures can be due to either relativistic jets in the
plane of the sky or mildly relativistic jets. For our BCGs, we
exclude the first hypothesis because of statistical
considerations and after comparison with the BCS results. It is not possible
that all BCGs in cool core clusters with resolved jets
are oriented on the plane of sky. We therefore conclude that BCGs in
cool core clusters have on the parsec scale mildly relativistic jets. 
This hypothesis seems confirmed also looking at the observed total 
arcsecond radio power at 408 MHz versus the observed arcsecond core radio power at 5 GHz
for all the sources in our extended sample: the two-sided sources detected in cooling clusters, which are expected
to show mildly relativistic jets, do not agree with the
correlation found by \cite{gio01}, while the BCG with one-sided jets are in good agreement with the general correlation and this confirms the presence of relativistic jets in these BCGs (\cite{liu10}).

\subsection{Mildly relativistic jets.}
\cite{ro08} discussed the interaction between relativistic jets and
the surrounding ISM. They showed that a jet perturbation grows because
of Kelvin-Helmotz instability and produces a strong interaction between the
jet and the external medium with a consequent mixing and
deceleration.  The deceleration becomes more efficient as the density
ratio of the ambient medium to the jet increases.

Because of the dense ISM of BCGs in cool
core clusters \cite{sal03}, we suggest that within BCGs in cool core clusters the jet
interaction with the ISM is already relevant on the parsec scale. We note also that two-sided jets are present only in BCGs at the center of clusters with a central mass accretion rate $>$ 90  M$_{\odot}$/yr \cite{liu10}.
In this
scenario, as for sources in non cool core clusters, the jet begin
relativistic (and thus one-sided at its base) but a large
value of the density ratio can produce a sub-relativistic (and therefore
two-sided) heavy jet at a much shorter distance from the central 
engine compared to ``normal'' FR I radio galaxies. This suggestion
is supported  by literature data on Hydra A and 3C 84 \cite{liu10}, BCGs of two cool
core clusters (A780 and A426 respectively).

\section{New 1.6 GHz VLBA observations on two peculiar BCGs. }
Among the undetected sources (Sect. \ref{res}), we found 2 peculiar cases of BGCs, IC 712 and NGC 6047 for which the non detection in our previous VLBA data at 5 GHz was not expected and not at all understood. In particular:

\begin{itemize}
\item {\bf IC 712} is the BGC of the main condensation of a merging cluster Abell 1314. This cluster is a binary
cluster with two main condensations visible in the X-ray images and it exhibits very clumpy, elongated X-ray emission, and a strong
X-ray centroid shift \cite{bli98}. The BCG of the second condensation is IC 708. IC\,712 has a WAT structure of very small linear
size ($\sim$ 4.6 kpc) with a total flux density at 1.4 GHz of 26.3 mJy
\cite{gio94}. 
In our VLBA maps 5 GHz, differently to IC 708 (one-sided at parsec scale), IC 712 is undetected at the 5$\sigma$ level ($<$0.45 mJy/
beam). 
\item {\bf NGC 6047} is a BGC of the second condensation of Abell 2151. NGC 6041A is the BCG with the first main condensation. This Abell cluster is also
known as the Hercules cluster. It is a highly structured cluster
despite the apparently regular velocity distribution of the main
field. This cluster should be considered as an ongoing cluster
merging. 
NGC 6047 is classified as an E/S0 and has an unusual optical morphology
indicative of a recent merger that severely
disrupted the dynamics of an HI disk whose gas lost its
kinematic coherence \cite{di97}. On the kiloparsec scale, NGC 6047 exhibits
an extended FRI structure with a two-sided jet emission
\cite{fe88}. The northern jet is brighter (flux density at 5 GHz
$\sim$13.6 mJy) and more collimated. The total flux density at 1.4 GHz
is $\sim$ 728 mJy and the core flux density at 5 GHz is $\sim$8.3
mJy. In our VLBA maps at 5 GHz, differently to NGC 6041A (unresolved on parsec scale), NGC 6047 appears undetected above 5$\sigma$= 0.55 mJy/beam.
\end{itemize}

We suggested that the non detection of these two sources is the result of complex
structure in intermediate scale that we have not able to map with the resolution of our previous
VLBA observations at 5 GHz. This possible structure could be due to a strong interaction of the
parsec scale jet and the ISM (InterStellar Medium) and could produce a non relativistic jets as
discussed for 4C 26.42 (Fig. 1, see also \cite{liu09}). The presence of sub-kpc emission has been suggested
in many sources (\cite{liu10}) to explain the large discrepancy between the parsec and kpc
core flux density. 

\begin{figure}
\centering
\includegraphics[width=0.7\textwidth]{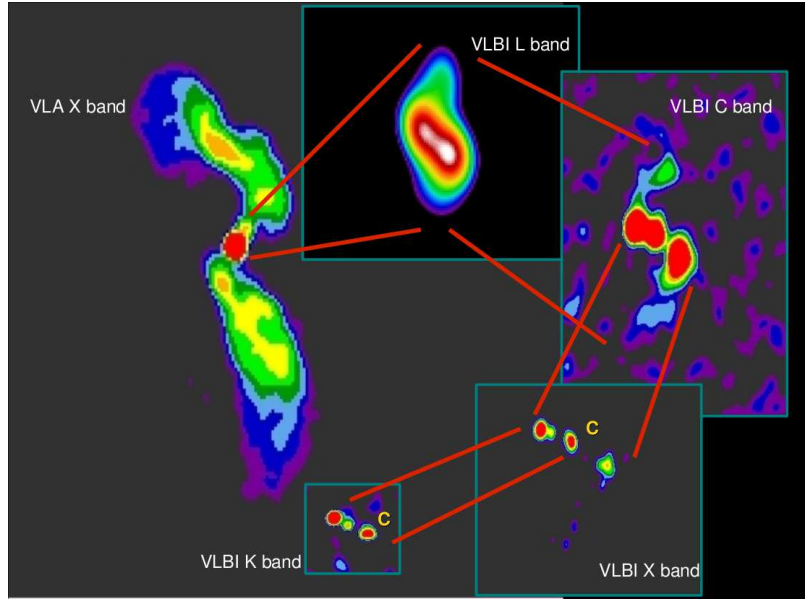}
\caption{Clockwise from left to right, zooming from kiloparsec to mas scale radiostructure of 4C 26.42 \cite{liu09}: color maps of VLA X band, VLBI L band, VLBI C band, VLBI X band and VLBI K band data. (C) indicates the core component.}
\end{figure}

For all these reasons, for both of these BCGs, we asked and obtained in July 2010 new VLBA observations at 1.6 GHz for a total number of 8 hours (4 hours for each target in phase reference mode considering also the time on calibrator sources). This allowed us an intermediate resolution and better sensitivity increased also by requesting a large recording rate (512 Mbps) and larger bandwidth (16 MHz).

The resolution achieved in these new VLBA data at 1.6 GHz is $\sim$16 $\times$ 11 mas. In both cases, the sources are again non detected above 5 $\sigma$ = 0.5 mJy/beam. Looking at clear kpc scale core emission at few mJy level, we excluded radio quiet core. Problem of wrong core coordinates were also excluded as for these sources core positions are available from VLA. Similarly to previous 5 GHz data, we imaged nevertheless large field of view ($\sim$4 arcsec) excluding possible source shift from the center pointing. We therefore suggest that the core emission for these source could be so diffuse that also these new VLBA observations are not able to map it. Future e-VLA observations at high frequency (e.g. 22 GHz) and/or deeper EVN data will be necessary to map properly the parsec core structured emission of these two peculiar BCGs.

\section{Acknowledgments.}
This work was supported by contributions of European Union, Valle D'Aosta Region and the Italian Minister for Work and Welfare. We thank the organizers of a very interesting meeting. The National Radio Astronomy Observatory is operated by Associated Universities, Inc., under cooperative agreement with the National Science Foundation.

\end{document}